\documentclass[sigconf]{acmart}
\acmConference[Work-in-progress]{}{}{}

\setcopyright{acmcopyright}
\copyrightyear{2018}
\acmYear{2018}
\acmDOI{10.1145/1122445.1122456}

\usepackage{tikz}
\usetikzlibrary{shapes,shapes.geometric,arrows.meta}
\tikzset{%
  >={Latex[width=2mm,length=2mm]},
            base/.style = {rectangle, rounded corners,
            draw=black,
                            minimum width=2.9cm, minimum height=1cm, 
                           text centered, font=\sffamily},
  activityStarts/.style = {base, fill=white!30},
       startstop/.style = {base, fill=white!30}, 
    activityRuns/.style = {base, fill=white!30}, 
         process/.style = {base, text width=2.5cm}, 
        db/.style = {cylinder, draw=black, 
        fill=white!30,
        shape border rotate=90, 
        draw,
        minimum height=2cm,
        minimum width=1.5cm,
        shape aspect=.25,
        text width = 2cm,
        text centered}
}

\usepackage{fancybox}
\usepackage[ textsize=tiny, textwidth=1.4cm]{todonotes}
\newcommand{\npa}{neural program analyzer\xspace}
\newcommand{\npas}{neural program analyzers\xspace}

\newcommand{\Npa}{Neural program analyzer\xspace}

\newcommand{\NPAS}{Neural Program Analyzers\xspace}

\newcommand{\iTODO}[1]{\todo[inline,color=orange!10,size=\small]{\textbf{TODO:} #1}}

\newcommand{\Space}[1]{}

\newcommand{\Etal}{\emph{et al.}}

\usepackage{xspace}

\newcommand{\Comment}[1]{}
\newcounter{observation}
\newcommand{\observation}[1]{\refstepcounter{observation}
        \begin{center}
        \Ovalbox{
        \begin{minipage}{0.93\columnwidth}
                \textbf{Observation \arabic{observation}:} #1
        \end{minipage}
        }
        \end{center}
}

\newcommand{\Part}[1]{\noindent\textbf{#1}}

\newcommand{\ctv}{\emph{code2vec}\xspace}
\newcommand{\cts}{\emph{code2seq}\xspace}

\newcommand{\VN}{\emph{Variable Renaming}\xspace}
\newcommand{\BX}{\emph{Boolean Exchange}\xspace}
\newcommand{\LX}{\emph{Loop Exchange}\xspace}
\newcommand{\PS}{\emph{Permute Statement}\xspace}
\newcommand{\SF}{\emph{Switch to If}\xspace}
\newcommand{\TC}{\emph{Try-Catch Insertion}\xspace}
\newcommand{\UN}{\emph{Unused Statement}\xspace}

\newcommand{\Js}{\textsc{Java-Small}\xspace}
\newcommand{\Jm}{\textsc{Java-Med}\xspace}
\newcommand{\Jl}{\textsc{Java-Large}\xspace}

\usepackage{graphicx}
\usepackage{float}
\usepackage{pbox}
\usepackage{listings}
\usepackage[labelfont=bf,textfont=md]{caption}
\usepackage{multirow}
\usepackage{xspace}

\usepackage{array,etoolbox}
\preto\tabular{\setcounter{magicrownumbers}{0}}
\newcounter{magicrownumbers}
\def\rownumber{}
\newcolumntype{Z}{>{\setbox0=\hbox\bgroup}c<{\egroup}@{\hspace*{-\tabcolsep}}}

\begin{document}

\title{Evaluation of Generalizability of Neural Program Analyzers under Semantic-Preserving Transformations}
\author{Md. Rafiqul Islam Rabin}
\affiliation{%
  \institution{University of Houston}
  \city{Houston, TX}
  \country{USA}}
\author{Mohammad Amin Alipour}
\affiliation{%
  \institution{University of Houston}
  \city{Houston, TX}
  \country{USA}}




\newcommand{\dypro}{\textsc{DyPro}\xspace}
\newcommand{\coset}{\textsc{CoSet}\xspace}

\begin{abstract}
The abundance of publicly available source code repositories, in conjunction with the advances in neural networks, has enabled data-driven approaches to program analysis. These approaches, called neural program analyzers, use neural networks to extract patterns in the programs for tasks ranging from development productivity to program reasoning. Despite the growing popularity of neural program analyzers, the extent to which their results are generalizable is unknown. 

In this paper, we perform a large-scale evaluation of the generalizability of two popular neural program analyzers using seven semantically-equivalent transformations of programs. Our results caution that in many cases the neural program analyzers fail to generalize well, sometimes to programs with negligible textual differences. The results provide the initial stepping stones for quantifying robustness in neural program analyzers.

\end{abstract}

\maketitle
\section{Introduction}

\newcommand{\eg}{\textit{e.g.}\xspace}
\newcommand{\ie}{\textit{i.e.}\xspace}
\newcommand{\etal}{\textit{et al.}\xspace}

Abundance of publicly available source code repositories has enabled a surge in data-driven approaches to programs analysis tasks. Those approaches aim to discover common programming patterns for various downstream applications\cite{BigCodeSurvey},
\eg, prediction of data types in dynamically typed languages~\cite{vincent:type}, detection of the variable naming issues~\cite{allamanis2017learning}, or repair of software defects~\cite{dinella2020hoppity}. 
The advent of deep neural networks has accelerated the innovation in this area and has greatly enhanced the performance of these approaches. 
The performance of deep neural networks in cognitive tasks such as method name prediction or variable naming has reached or exceeded the performance of other data-driven approaches. 
The performance of neural networks has encouraged researchers to increasingly adopt the neural networks in the program analysis giving rise to increasing use of neural program analyzers.

While the performance of the neural program analyzers
continues to improve, the extent to which they can generalize to new, unseen programs is still unknown. 
This problem is of more importance if we want to use them in downstream safety-critical tasks, such as malware detection.
This problem is particularly hard, as the interpretation of neural models that constitute the core reasoning engine of \npas remains to be challenging---especially for the complex neural networks (\eg, RNN) that are commonly used for processing source code.


Reliable application of neural program analyzers requires awareness of the limits of these analyzers.
A complete understanding of the extent of usefulness of such approaches would help developers to know when to use data-driven approaches and when to resort to traditional deductive methods of program analysis.
It also would help researchers to focus their efforts on devising techniques to alleviate the shortcomings of these analyzers.
Lack of knowledge about the limits of the neural program analyzers may exaggerate their capability and cause careless applications of the analyzers on the domains that they are not suited for; or, spending time and efforts on developing neural program analyzers while a traditional, more understandable technique can perform equally well or better.

Recently, we have seen a growing interest in the rigorous evaluation of neural program analyzers. ~\citet{Wang:2019:Coset} compared the robustness of different program representation under compiler optimization transformations. They found that the program representations based on static code features are more sensitive to such changes than dynamic code features.~\citet{Milto:Onward:2019} evaluated the impact of code duplication in various neural program analyzers and found that code duplication in the training and test datasets inflated the performance of almost all current \npas. More recently, more preliminary studies in this field started to emerge; \eg, ~\citet{Rabin:ASE:2019a} proposed the idea of testing \npas using semantic-preserving transformations, and ~\citet{Yefet:Arxiv:2019} followed and proposed adversarial example generation for \npas using prediction attribution \cite{Attribution}. 



\begin{figure*}[t!]
\noindent \begin{minipage}{.5\textwidth}
\begin{lstlisting}[title= Prediction before transformation: \textbf{\textcolor{blue}{compareTo}}]
public int compareTo(ApplicationAttemptId <@\textbf{\textcolor{blue}{other}}@>) {
    int compareAppIds = this.getApplicationId()
        .compareTo(<@\textbf{\textcolor{blue}{other}}@>.getApplicationId());
    if (compareAppIds == 0) {
        return this.getAttemptId() - <@\textbf{\textcolor{blue}{other}}@>.getAttemptId();
    } else {
        return compareAppIds;
    }
}
\end{lstlisting}
\end{minipage}%
\begin{minipage}{.5\textwidth}
\begin{lstlisting}[title= Prediction after transformation: \textbf{\textcolor{red}{getCount}}]
public int compareTo(ApplicationAttemptId <@\textbf{\textcolor{red}{var0}}@>) {
    int compareAppIds = this.getApplicationId()
        .compareTo(<@\textbf{\textcolor{red}{var0}}@>.getApplicationId());
    if (compareAppIds == 0) {
        return this.getAttemptId() - <@\textbf{\textcolor{red}{var0}}@>.getAttemptId();
    } else {
        return compareAppIds;
    }
}
\end{lstlisting}
\end{minipage}
\caption{\VN on \texttt{java-small/test/hadoop/ApplicationAttemptId.java} file.}
\label{fig:motivating-example}
\end{figure*}

\Part{Goal:}
In this paper, we attempt to understand the limits of \emph{generalizability} of neural program analyzers by comparing their behavior under semantic-preserving transformations; that is, how the result of a \npa generalizes to a semantically-equivalent program.
We should note that intent from generalizability differs from Kang \etal ~\cite{Kang:ASE:2019:Generalizability}. 
Kang \etal, in fact, evaluate the usefulness of a \npa in downstream tasks, while we evaluate their generalizability to semantically-equivalent programs.
Moreover, this work, while related to, does not address the neural robustness~\cite{szegedy2013intriguing} in \npas, as robustness requires imperceptible changes to programs. 
Although the impact of our transformations on the semantics of the program is imperceptible, the change on the textual structure of the program is not necessarily imperceptible. Nonetheless, this work can be considered as a stepping stone towards the robustness of \npas.

In this paper, we report the results of a study on the generalizability of two highly-cited neural program analyzers: code2vec~\cite{alon2018code2vec} and code2seq~\cite{alon2018code2seq}. 
In this study, we transform programs from the original dataset code2seq is trained on to generate semantically-equivalent counterparts. 
We employ seven semantic-preserving transformations that impact the structure of programs (\ie abstract syntax tree) with varying degrees.

Our results suggest that the models evaluated in this study are sensitive to the transformation and in many cases, transformations of a program lead the \npas to produce a different prediction than the \npas would produce on the original program. This sensitivity remains an issue even in the case of very small changes to the programs, such as renaming variables or reordering independent statements in a basic block. 
The result of this study is a cautionary tale that reveals that the generalizability of \npas is still far from ideal and require more attention from the research community to devise robust models of code and canonicalized program representation. 

\Part{Contributions.} This paper makes the following contributions.
\begin{itemize}
\item We introduce the notion of generalizability in neural program analyzers.
\item We perform a large-scale study to evaluate the state-of-the-art \npas.
\item We discuss the practical implication of our results.
\end{itemize}

\section{Motivating Example}

We use code2vec~\cite{alon2018code2vec} for exposition in this section.
The code2vec~\cite{alon2018code2vec} is a \npa that predicts the name of a Java method given its body. Such \npa, in addition to other developer productivity tools, can be useful in classification of code, and code similarity detection.

Figure ~\ref{fig:motivating-example} shows two semantically-identical functions that implement \texttt{compareTo} function. The only difference between them is in the name of one of the variables. The snippet on the left uses \texttt{other}, while the code on the right uses \texttt{var0}. The results of code2vec however on these semantically equivalent programs are drastically different. code2vec predicts the snippet on the left to be \texttt{compareTo} function, and the function on the right to be \texttt{getCount}.
It seems that code2vec heavily relied on the variable name \texttt{other} for its correct prediction.

Lack of robustness to modest changes hampers the wider application of \npas beyond developer productivity tasks, particularly in the mission-critical problem settings where higher levels of robustness and generalizability are required
\eg, malware detection.
Despite the significant progresses made in novel application of neural networks for program analysis tasks, their generalizability with respect to program transformations have not been adequately explored. 

\section{Background}
Most \npas are essentially classifiers that take a code snippet or a whole program as an input, and make predictions about some of its characteristics; \eg, a bug prediction classifier that predicts the buggy-ness of statements in the input program.

Performance of a \npa mainly depends on three main components: quality of data, the neural network architecture and its learning parameters, and the representation of data for the neural network.

Currently, most studies use open-source projects usually in mainstream programming languages, \eg, C\#, Java, C, or JavaScript. 
The available standard datasets for these tasks are still very young and their quality is somewhat unknown.
For example, a recent study by Allamanis~\cite{Milto:Onward:2019} showed that virtually all available datasets suffer from code duplication that can greatly impact the performance of \npas.

Another factor in the performance of \npas is source code representation. Since neural networks can only take vectors of numbers, source code embeddings are used to produce a vector representation of source code.
The representation determines which program features and how to be represented in the vector embeddings. The representations can be broadly categorized into two categories: static and dynamic. Static program representations consider only the features that can be extracted from parsing text of the programs, while dynamic representations include some features pertaining to the real execution of programs. 

The third building block in building a \npa is a neural network architecture and learning parameters. There are numerous choices of network architectures each with different properties.
It seems that the class of recurrent neural networks (\eg, LSTM) and graph neural networks are among the most popular architecture in \npas.

\section{Evaluation Approach}
\begin{figure}
	\centering
	\begin{tikzpicture}[scale=0.8, transform shape,
	    auto, node distance=3.3cm,
	]
	\node (orig)             [db]              {Original Programs};
	\node (transform)     [process, right of=orig]          {Transformation Engine};
	\node (transformed)      [db, right of=transform]   {Transformed Programs};
	\node (model)     [process, below of=transform, yshift=+1.5cm]   {\Npa};
	\draw[->]             (orig) -- (transform);
	\draw[->]             (transform) -- (transformed);
	\draw[->]             (orig) -- (model);
	\draw[->]             (transformed) -- (model);
	\end{tikzpicture}%
	\caption{The workflow of evaluation in this study.}
	\label{fig:workflow}
\end{figure}
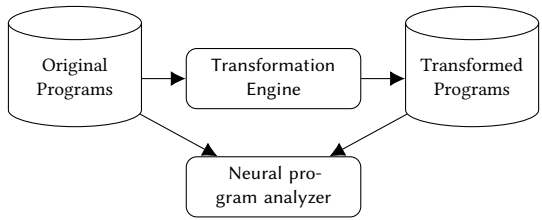

In this section, we describe the experimental setting and our evaluation approach.
Figure~\ref{fig:workflow} depicts an overall view of the evaluation process.
Our approach relies on metamorphic relations that state the output of a \npa should not substantially differ on semantically-equivalent programs. 
It is similar to the notion of local fidelity of classifiers that state that the behavior of classifiers should not change substantially in the vicinity of an input~\cite{WhyShouldITrustYou}. 
The approach can broadly be divided into two main steps: (1) semantic-preserving program transformation, and (2) discrepancy identification (Oracle). 

\subsection{Transformations}
We have used the following seven transformations to generate semantically-equivalent programs.

\begin{itemize}
    \item  \textbf{\VN (VN)} renames the name of a variable. The new name of the variable will be in the form of \texttt{varN} for a value of N such that \texttt{N} that has not been defined in the scope.
    \item \textbf{Loop Exchange (LX)} replaces \texttt{for} loops with \texttt{while} loops or vice versa.
    \item \textbf{\SF (SF)} replaces a \texttt{switch} statement in the program with its equivalent \texttt{if} statement.
    \item \textbf{Boolean Exchange (BX)} switches the value of a boolean variable from \texttt{true} to \texttt{false} or vice versa, and propagates this change in the program to ensure a semantic equivalence of the transformed program with the original program. 
    \item \textbf{\PS (PS)} swaps two independent statements (i.e., with no dependence) in a basic block.  
    \item \textbf{\TC (TC)} adds \texttt{try-catch} statements to a basic block. The \texttt{catch} captures the highest exception in Java, \ie \texttt{Exception}. Note that although this transformation does not change the behavior of the program in the normal executions, it alters the error-handling behavior of the program.  
    \item \textbf{\UN (UN)} inserts an unused string declaration to a random basic block in the program.
\end{itemize}

Note that each transformation has a different impact on the structure of programs:

\begin{itemize}
    \item The \VN transformation only changes the terminal values and does not affect the structure of AST.
    \item The \BX transformation alters the value of \texttt{true} or \texttt{false} and modifies the structure of AST by removing or inserting unary-not nodes.
    \item The \LX transformation extensively impacts the AST by removing and inserting nodes.
    \item The \SF also impacts the AST of the program substantially by removing and inserting nodes. 
    \item The \PS transformation does not change the nodes in AST, rather it only reorders two subtrees in the AST.
    \item The \TC transformation modifies the structure of AST by adding additional nodes and branches to realize the \texttt{try-catch} block.
    \item The \UN transformation adds a few nodes into the AST which increases the number of paths.
\end{itemize}

\subsection{Oracle}
Given the original program and the transformed, we try both in the \npa and compare the results.
We compare the predicted label of the analyzer in both original and transformed programs.
Ideally, the neural model should behave similarly with both the original and the transformed program. 
A discrepancy between the predicted label on the original program and the transformed program is considered prediction change.



\section{Experimental Setting}

In this section, we describe the models and datasets that we used in the evaluation.

\subsection{Subject \NPAS}

We have used \ctv ~\cite{alon2018code2vec} and \cts  ~\cite{alon2018code2seq} \npas for the experimentation.
Their task is to predict a method's name given the body of a method.

\ctv uses a bag of AST paths to model the source code.
Each path consists of a pair of terminals in the abstract syntax tree and their corresponding path between them in the AST.
The path, along with source and destination terminals are mapped into its vector embeddings which are learned jointly with other network parameters during training.
The three separate vectors of each path-context are then concatenated to a single context vector using a fully connected layer which is learned during training with the network.
An attention vector is also learned with the network which is used to score each path-context and aggregate multiple path-contexts to a single code vector representing the method's body.
After that, the model predicts the probability of each target method's name given the code vector of method's body with a softmax-normalization between the code vector and each of the embeddings of target method's name.

\ctv uses monolithic path embeddings and only generates a single label at a time, while the \cts model uses an encoder-decoder architecture to encode paths node-by-node and generate label as sequences at each step.
The encoder represents a method's body as a set of AST paths where each path is compressed to a fixed-length vector using a bi-directional LSTM which encodes paths node-by-node.
The decoder uses attention to select relevant paths while decoding and predicts sub-tokens of target sequence at each step when generating the method's name.

\subsection{Datasets}
The datasets published along with \ctv only contains the programs in preprocessed format, but, for this study, we needed the raw Java files to make the transformations.
Fortunately, the datasets accompanied by \cts contained the java files, and \cts and \ctv shared the same code pre-processing techniques. Therefore, we used the \cts dataset for training \npas for the study.


There are three Java datasets:\Js, \Jm, and \Jl.
\begin{itemize}
\item \Js: This dataset contains 9 Java projects for training, 1 for validation and 1 for testing. Overall, it contains about 700K methods. The compressed size is about 366MB and the extracted size is about 1.9GB.
\item \Jm: This dataset contains 800 Java projects for training, 100 for validation and 100 for testing. Overall, it contains about 4M examples. The compressed size is about 1.8GB and the extracted size is about 9.3GB.
\item \Jl: This dataset contains 9000 Java projects for training, 200 for validation and 300 for testing. Overall, it contains about 16M examples. The compressed size is about 7.2GB and the extracted size is about 37GB.
\end{itemize}

\subsection{Training Models per Datasets}
The authors of \ctv and \cts have made the programs for training a neural model and evaluating it public. We use their tools on three aforementioned datasets to train and create three \ctv \npas  and three \cts \cts.


We train each model up to 100 epochs and save after each epoch. 
We stop in an earlier epoch if the F1 score, in that epoch, is reasonably close to the score reported in the corresponding papers. Otherwise, we continue for 100 epochs and select the model in the epoch with the highest F1 score.
Table~\ref{table:all_models} summarizes the characteristics of the trained models.
While the performance of our trained models for \cts is on par with to the ones reported in the corresponding paper~\cite{alon2018code2seq}, the performance of \ctv, after 100 epochs did not reach the performance reported in \cite{alon2018code2vec}, perhaps due to the differences in the dataset. 
However, the performance of our trained \ctv models is similar to the one reported in  ~\cite{alon2018code2seq}.

\begin{table}
    \begin{center}
        \caption{Characteristics of the models in \npa.}
        \label{table:all_models}
        \resizebox{\columnwidth}{!}{%
        \begin{tabular}{|c|c|c|c|c|c|}
            \hline
            \textbf{Model} & \textbf{Dataset} &
            \textbf{Precision} & \textbf{Recall} & \textbf{F1 Score} \\ \hline 
            \hline
            \multirow{3}{*}{\ctv} & \Js & 28.36 & 22.37 & 25.01 \\ \cline{2-5}
                                  & \Jm   & 42.55 & 30.85 & 35.76 \\ \cline{2-5}
                                  & \Jl & 45.17 & 32.28 & 37.65 \\ \hline
            \hline 
            \multirow{3}{*}{\cts} & \Js & 46.30 & 38.81 & 42.23 \\ \cline{2-5}
                                  & \Jm   & 59.94 & 48.03 & 53.33 \\ \cline{2-5}
                                  & \Jl & 64.03 & 55.02 & 59.19 \\ \hline
        \end{tabular}%
        }
    \end{center}
\end{table}

\subsection{Population of Transformed Programs}
We apply the applicable transformations to the program in the testing data of the above-mentioned datasets. The number of original programs in our study is 2,088,411 and we used transformation to create 4,075,949 transformed programs.
The types and number of applicable transformations vary from a program to another.
Therefore, in our approach, different methods, based on the language features that they use, produce a different number of transformed programs.

Overall, the number of original programs with incorrect predictions is, on average, 2.4 times higher than the number of programs with correct predictions. Moreover, programs with incorrect predictions are amenable to, on average, 1.4 times more transformations. It may suggest that programs with correct predictions are smaller and simpler. 
In total, the number of transformed programs from the program with wrong initial predictions is much higher (3.4x and higher) than the number of transformed programs from programs with correct initial predictions.

\subsection{Research Questions}
In this paper, we seek to answer the following research questions.
\begin{itemize}
    \item[RQ1] How do the transformations impact the prediction of \npas?
    \item[RQ2] When transformations are most effective in modifying the prediction of the \npas?
    \item[RQ3] How does method length impact the \npa's generalizability?
\end{itemize}


\section{Results}

\subsection{RQ1: Impact of Transformation on the Models}
\begin{table}
    \caption{Change of prediction for \ctv on \Js dataset.}
    \label{table:code2vec_js}
    \begin{tabular}{|c|c|c|}
        \hline
        \textbf{Transformation} 
        & \textbf{\#} 
        & \textbf{Total} 
        
        \\  \hline \hline

        \multirow{4}{*}{\VN} & \# Original programs & 31113 \\  \cline{2-3}
          & \# Transformed programs & 123123 \\  \cline{2-3}
          & \# Prediction changing programs & 67622 \\  \cline{2-3}
          & Prediction change(\%) & 54.92 \\  \hline
        \hline
        
        \multirow{4}{*}{\BX} & \# Original programs & 1158 \\  \cline{2-3}
          & \# Transformed programs & 1519 \\  \cline{2-3}
          & \# Prediction changing programs & 818 \\  \cline{2-3}
          & Prediction change(\%) & 53.85 \\  \hline
        \hline
        
        \multirow{4}{*}{\LX} & \# Original programs & 3699 \\  \cline{2-3}
          & \# Transformed programs & 5160 \\  \cline{2-3}
          & \# Prediction changing programs & 3064 \\  \cline{2-3}
          & Prediction change(\%) & 59.38 \\  \hline
        \hline
        
        \multirow{4}{*}{\SF} & \# Original programs & 246 \\  \cline{2-3}
          & \# Transformed programs & 259 \\  \cline{2-3}
          & \# Prediction changing programs & 178 \\  \cline{2-3}
          & Prediction change(\%) & 68.73 \\  \hline
        \hline
        
        \multirow{4}{*}{\PS} & \# Original programs & 15325 \\  \cline{2-3}
          & \# Transformed programs & 74950 \\  \cline{2-3}
          & \# Prediction changing programs & 53791 \\  \cline{2-3}
          & Prediction change(\%) & 71.77 \\  \hline
        \hline
        
        \multirow{4}{*}{\TC} & \# Original programs & 32078 \\  \cline{2-3}
          & \# Transformed programs & 32078 \\  \cline{2-3}
          & \# Prediction changing programs & 15039 \\  \cline{2-3}
          & Prediction change(\%) & 46.88 \\  \hline
        \hline
        
        \multirow{4}{*}{\UN} & \# Original programs & 44426 \\  \cline{2-3}
          & \# Transformed programs & 44426 \\  \cline{2-3}
          & \# Prediction changing programs & 17755 \\  \cline{2-3}
          & Prediction change(\%) & 39.97 \\  \hline
        
    \end{tabular}
\end{table}

\begin{table}
    \caption{Change of prediction for \ctv on \Jm dataset.}
    \label{table:code2vec_jm}
    \begin{tabular}{|c|c|c|}
        \hline
        \textbf{Transformation} 
        & \textbf{\#} 
        & \textbf{Total} 
        
        \\  \hline \hline
        
        \multirow{4}{*}{\VN} & \# Original programs & 235961 \\  \cline{2-3}
          & \# Transformed programs & 771208 \\  \cline{2-3}
          & \# Prediction changing programs & 358984 \\  \cline{2-3}
          & Prediction change(\%) & 46.55 \\  \hline
        \hline
        
        \multirow{4}{*}{\BX} & \# Original programs & 6407 \\  \cline{2-3}
          & \# Transformed programs & 8840 \\  \cline{2-3}
          & \# Prediction changing programs & 4451 \\  \cline{2-3}
          & Prediction change(\%) & 50.35 \\  \hline
        \hline
        
        \multirow{4}{*}{\LX} & \# Original programs & 17107 \\  \cline{2-3}
          & \# Transformed programs & 23533 \\  \cline{2-3}
          & \# Prediction changing programs & 14772 \\  \cline{2-3}
          & Prediction change(\%) & 62.77 \\  \hline
        \hline
        
        \multirow{4}{*}{\SF} & \# Original programs & 3312 \\  \cline{2-3}
          & \# Transformed programs & 3839 \\  \cline{2-3}
          & \# Prediction changing programs & 2300 \\  \cline{2-3}
          & Prediction change(\%) & 59.91 \\  \hline
        \hline
        
        \multirow{4}{*}{\PS} & \# Original programs & 88865 \\  \cline{2-3}
          & \# Transformed programs & 366840 \\  \cline{2-3}
          & \# Prediction changing programs & 232054 \\  \cline{2-3}
          & Prediction change(\%) & 63.26 \\  \hline
        \hline
        
        \multirow{4}{*}{\TC} & \# Original programs & 232769 \\  \cline{2-3}
          & \# Transformed programs & 232769 \\  \cline{2-3}
          & \# Prediction changing programs & 99878 \\  \cline{2-3}
          & Prediction change(\%) & 42.91 \\  \hline
        \hline
        
        \multirow{4}{*}{\UN} & \# Original programs & 351621 \\  \cline{2-3}
          & \# Transformed programs & 351621 \\  \cline{2-3}
          & \# Prediction changing programs & 125880 \\  \cline{2-3}
          & Prediction change(\%) & 35.8 \\  \hline

    \end{tabular}
\end{table}

\begin{table}
    \caption{Change of prediction for \ctv on \Jl dataset.}
    \label{table:code2vec_jl}
    \begin{tabular}{|c|c|c|}
        \hline
        \textbf{Transformation} 
        & \textbf{\#} 
        & \textbf{Total} 
        
        \\  \hline \hline

        \multirow{4}{*}{\VN} & \# Original programs & 252725 \\  \cline{2-3}
          & \# Transformed programs & 916565 \\  \cline{2-3}
          & \# Prediction changing programs & 385466 \\  \cline{2-3}
          & Prediction change(\%) & 42.06 \\  \hline
        \hline
        
        \multirow{4}{*}{\BX} & \# Original programs & 8868 \\  \cline{2-3}
          & \# Transformed programs & 12107 \\  \cline{2-3}
          & \# Prediction changing programs & 5787 \\  \cline{2-3}
          & Prediction change(\%) & 47.8 \\  \hline
        \hline
        
        \multirow{4}{*}{\LX} & \# Original programs & 35565 \\  \cline{2-3}
          & \# Transformed programs & 49665 \\  \cline{2-3}
          & \# Prediction changing programs & 23104 \\  \cline{2-3}
          & Prediction change(\%) & 46.52 \\  \hline
        \hline
        
        \multirow{4}{*}{\SF} & \# Original programs & 10478 \\  \cline{2-3}
          & \# Transformed programs & 11165 \\  \cline{2-3}
          & \# Prediction changing programs & 3386 \\  \cline{2-3}
          & Prediction change(\%) & 30.33 \\  \hline
        \hline
        
        \multirow{4}{*}{\PS} & \# Original programs & 98669 \\  \cline{2-3}
          & \# Transformed programs & 428263 \\  \cline{2-3}
          & \# Prediction changing programs & 243574 \\  \cline{2-3}
          & Prediction change(\%) & 56.87 \\  \hline
        \hline
        
        \multirow{4}{*}{\TC} & \# Original programs & 247092 \\  \cline{2-3}
          & \# Transformed programs & 247092 \\  \cline{2-3}
          & \# Prediction changing programs & 88609 \\  \cline{2-3}
          & Prediction change(\%) & 35.86 \\  \hline
        \hline
        
        \multirow{4}{*}{\UN} & \# Original programs & 370927 \\  \cline{2-3}
          & \# Transformed programs & 370927 \\  \cline{2-3}
          & \# Prediction changing programs & 115781 \\  \cline{2-3}
          & Prediction change(\%) & 31.21 \\  \hline

    \end{tabular}
\end{table}

\begin{table}
    \caption{Change of prediction for \cts on \Js dataset.}
    \label{table:code2seq_js}
    \begin{tabular}{|c|c|c|}
        \hline
        \textbf{Transformation} 
        & \textbf{\#} 
        & \textbf{Total} 
        
        \\  \hline \hline

        \multirow{4}{*}{\VN} & \# Original programs & 31113 \\  \cline{2-3}
          & \# Transformed programs & 123123 \\  \cline{2-3}
          & \# Prediction changing programs & 70371 \\  \cline{2-3}
          & Prediction change(\%) & 57.16 \\  \hline
        \hline
        
        \multirow{4}{*}{\BX} & \# Original programs & 1158 \\  \cline{2-3}
          & \# Transformed programs & 1519 \\  \cline{2-3}
          & \# Prediction changing programs & 825 \\  \cline{2-3}
          & Prediction change(\%) & 54.31 \\  \hline
        \hline
        
        \multirow{4}{*}{\LX} & \# Original programs & 3699 \\  \cline{2-3}
          & \# Transformed programs & 5160 \\  \cline{2-3}
          & \# Prediction changing programs & 2711 \\  \cline{2-3}
          & Prediction change(\%) & 52.54 \\  \hline
        \hline
        
        \multirow{4}{*}{\SF} & \# Original programs & 246 \\  \cline{2-3}
          & \# Transformed programs & 259 \\  \cline{2-3}
          & \# Prediction changing programs & 160 \\  \cline{2-3}
          & Prediction change(\%) & 61.78 \\  \hline
        \hline
        
        \multirow{4}{*}{\PS} & \# Original programs & 15325 \\  \cline{2-3}
          & \# Transformed programs & 74950 \\  \cline{2-3}
          & \# Prediction changing programs & 42685 \\  \cline{2-3}
          & Prediction change(\%) & 56.95 \\  \hline
        \hline
        
        \multirow{4}{*}{\TC} & \# Original programs & 32078 \\  \cline{2-3}
          & \# Transformed programs & 32078 \\  \cline{2-3}
          & \# Prediction changing programs & 15490 \\  \cline{2-3}
          & Prediction change(\%) & 48.29 \\  \hline
        \hline
        
        \multirow{4}{*}{\UN} & \# Original programs & 44426 \\  \cline{2-3}
          & \# Transformed programs & 44426 \\  \cline{2-3}
          & \# Prediction changing programs & 20257 \\  \cline{2-3}
          & Prediction change(\%) & 45.6 \\  \hline

    \end{tabular}
\end{table}

\begin{table}
    \caption{Change of prediction for \cts on \Jm dataset.}
    \label{table:code2seq_jm}
    \begin{tabular}{|c|c|c|}
        \hline
        \textbf{Transformation} 
        & \textbf{\#} 
        & \textbf{Total} 
        
        \\  \hline \hline
        
        \multirow{4}{*}{\VN} & \# Original programs & 235961 \\  \cline{2-3}
          & \# Transformed programs & 771208 \\  \cline{2-3}
          & \# Prediction changing programs & 375939 \\  \cline{2-3}
          & Prediction change(\%) & 48.75 \\  \hline
        \hline
        
        \multirow{4}{*}{\BX} & \# Original programs & 6407 \\  \cline{2-3}
          & \# Transformed programs & 8840 \\  \cline{2-3}
          & \# Prediction changing programs & 3952 \\  \cline{2-3}
          & Prediction change(\%) & 44.71 \\  \hline
        \hline
        
        \multirow{4}{*}{\LX} & \# Original programs & 17107 \\  \cline{2-3}
          & \# Transformed programs & 23533 \\  \cline{2-3}
          & \# Prediction changing programs & 10659 \\  \cline{2-3}
          & Prediction change(\%) & 45.29 \\  \hline
        \hline
        
        \multirow{4}{*}{\SF} & \# Original programs & 3312 \\  \cline{2-3}
          & \# Transformed programs & 3839 \\  \cline{2-3}
          & \# Prediction changing programs & 1597 \\  \cline{2-3}
          & Prediction change(\%) & 41.6 \\  \hline
        \hline
        
        \multirow{4}{*}{\PS} & \# Original programs & 88865 \\  \cline{2-3}
          & \# Transformed programs & 366840 \\  \cline{2-3}
          & \# Prediction changing programs & 170619 \\  \cline{2-3}
          & Prediction change(\%) & 46.51 \\  \hline
        \hline
        
        \multirow{4}{*}{\TC} & \# Original programs & 232769 \\  \cline{2-3}
          & \# Transformed programs & 232769 \\  \cline{2-3}
          & \# Prediction changing programs & 91176 \\  \cline{2-3}
          & Prediction change(\%) & 39.17 \\  \hline
        \hline
        
        \multirow{4}{*}{\UN} & \# Original programs & 351621 \\  \cline{2-3}
          & \# Transformed programs & 351621 \\  \cline{2-3}
          & \# Prediction changing programs & 141511 \\  \cline{2-3}
          & Prediction change(\%) & 40.25 \\  \hline

    \end{tabular}
\end{table}
 
\begin{table}
    \caption{Change of prediction for \cts on \Jl dataset.}
    \label{table:code2seq_jl}
    \begin{tabular}{|c|c|c|}
        \hline
        \textbf{Transformation} 
        & \textbf{\#} 
        & \textbf{Total} 
        
        \\  \hline \hline

        \multirow{4}{*}{\VN} & \# Original programs & 252725 \\  \cline{2-3}
          & \# Transformed programs & 916565 \\  \cline{2-3}
          & \# Prediction changing programs & 431131 \\  \cline{2-3}
          & Prediction change(\%) & 47.04 \\  \hline
        \hline
        
        \multirow{4}{*}{\BX} & \# Original programs & 8868 \\  \cline{2-3}
          & \# Transformed programs & 12107 \\  \cline{2-3}
          & \# Prediction changing programs & 6227 \\  \cline{2-3}
          & Prediction change(\%) & 51.43 \\  \hline
        \hline
        
        \multirow{4}{*}{\LX} & \# Original programs & 35565 \\  \cline{2-3}
          & \# Transformed programs & 49665 \\  \cline{2-3}
          & \# Prediction changing programs & 21112 \\  \cline{2-3}
          & Prediction change(\%) & 42.51 \\  \hline
        \hline
        
        \multirow{4}{*}{\SF} & \# Original programs & 10478 \\  \cline{2-3}
          & \# Transformed programs & 11165 \\  \cline{2-3}
          & \# Prediction changing programs & 3247 \\  \cline{2-3}
          & Prediction change(\%) & 29.08 \\  \hline
        \hline
        
        \multirow{4}{*}{\PS} & \# Original programs & 98669 \\  \cline{2-3}
          & \# Transformed programs & 428263 \\  \cline{2-3}
          & \# Prediction changing programs & 186411 \\  \cline{2-3}
          & Prediction change(\%) & 43.53 \\  \hline
        \hline
        
        \multirow{4}{*}{\TC} & \# Original programs & 247092 \\  \cline{2-3}
          & \# Transformed programs & 247092 \\  \cline{2-3}
          & \# Prediction changing programs & 87357 \\  \cline{2-3}
          & Prediction change(\%) & 35.35 \\  \hline
        \hline
        
        \multirow{4}{*}{\UN} & \# Original programs & 370927 \\  \cline{2-3}
          & \# Transformed programs & 370927 \\  \cline{2-3}
          & \# Prediction changing programs & 138865 \\  \cline{2-3}
          & Prediction change(\%) & 37.44 \\  \hline

    \end{tabular}
\end{table}

Tables~\ref{table:code2vec_js}--\ref{table:code2seq_jl} show the changes of prediction for each transformation in all \ctv and \cts \npas that we trained. 
Note that since the inputs to \ctv and \cts are body of methods, we use terms methods and programs interchangeably, in this section.
For each transformation, ``\# Original programs'' denotes the number of programs eligible for the transformation, ``\# Transformed programs'' denotes the number of transformed programs. Note that  \# Transformed programs can be larger than \# Original programs, as a program may have more than one place where the transformation is applicable.
``\# Prediction changing programs'' provides raw number of transformed programs that the prediction of \npa on the original and transformed program differ. 
``Prediction change(\%)'' denotes the percentage of transformed program that changed the output of the \npa.

\ctv is most sensitive to \PS transformation on all datasets. On the other hand, the \cts is most vulnerable to \SF, \VN, and \BX transformation on \Js, \Jm, and \Jl dataset, respectively.
\VN changes the text of terminals that change the embedding of path-context as well. 
However, \TC and \UN add some additional nodes and paths in AST. If models give less attention to those new paths, then the change is less effective.
The \UN and \TC keep the existing AST mostly intact, consequently, in most cases, the average percentage of changes in prediction of \UN and \TC is comparatively less than other transformations.

"Total" column in Tables~\ref{table:code2vec_js}--\ref{table:code2seq_jl} supports that \PS is more powerful than \VN in \ctv model whereas \VN is more effective than \PS to \cts model on all datasets.
The real-value embeddings of paths are different for \ctv and \cts.
In \ctv, an embedding matrix is initialized randomly for paths and learned during training, that contains rows that are mapped to each of the AST paths.
On the other hand, in \cts, each node of path comes from a learned embedding matrix and then a bi-directional LSTM is used to encode each of the AST paths separately.
The bi-directional LSTM reads the path once from beginning to the end (as original order) and once from end to beginning (in reverse order).
Therefore, the order change by \PS becomes less sensitive in \cts than \ctv.

Additionally, the \LX seems more effective than \SF on \Jm and \Jl dataset, but turned opposite on \Js dataset, for all models.
Moreover, the \BX shows poor performance on \Js and \Jm dataset, but becomes comparatively more effective on \Jl dataset, for all models. 
Another observation is that almost in all cases, the percentage of prediction change for all transformations are higher on \Js dataset, and then significantly drops on \Jm and \Jl dataset, respectively.

\observation{\ctv is most sensitive to \PS transformation on all datasets. On the other hand, the \cts is most vulnerable to \SF, \VN, and \BX transformation on \Js, \Jm, and \Jl dataset, respectively.}

\subsection{RQ2: When Transformations are most Effective?}

\subsubsection{Single place transformation vs All place transformation}
In our analysis, thus far, if a program has multiple candidates for a transformation, say $n$ candidates, for transformation, we only apply them one at the time and end up with $n$ distinct programs. We call this \emph{single-place} transformation.
Alternatively, we can replace \emph{all} the candidate in a single transformation, and end up with one transformed program for each transformation per program. 
We call this \emph{all-place} transformation.
In the next experiment, we evaluate the generalizability of \npas under all-place transformation for the following transformations: \LX, \VN, \BX, and \SF.
Note that the all-place transformation is not applicable to \PS, \TC, and \UN transformations, as we apply the \PS transformation on a pair of statements and the \TC and \UN on a randomly selected statement.

\begin{figure*}
\noindent \begin{minipage}{.5\textwidth}
\caption*{(a) \ctv}
\includegraphics[width=0.99\linewidth]{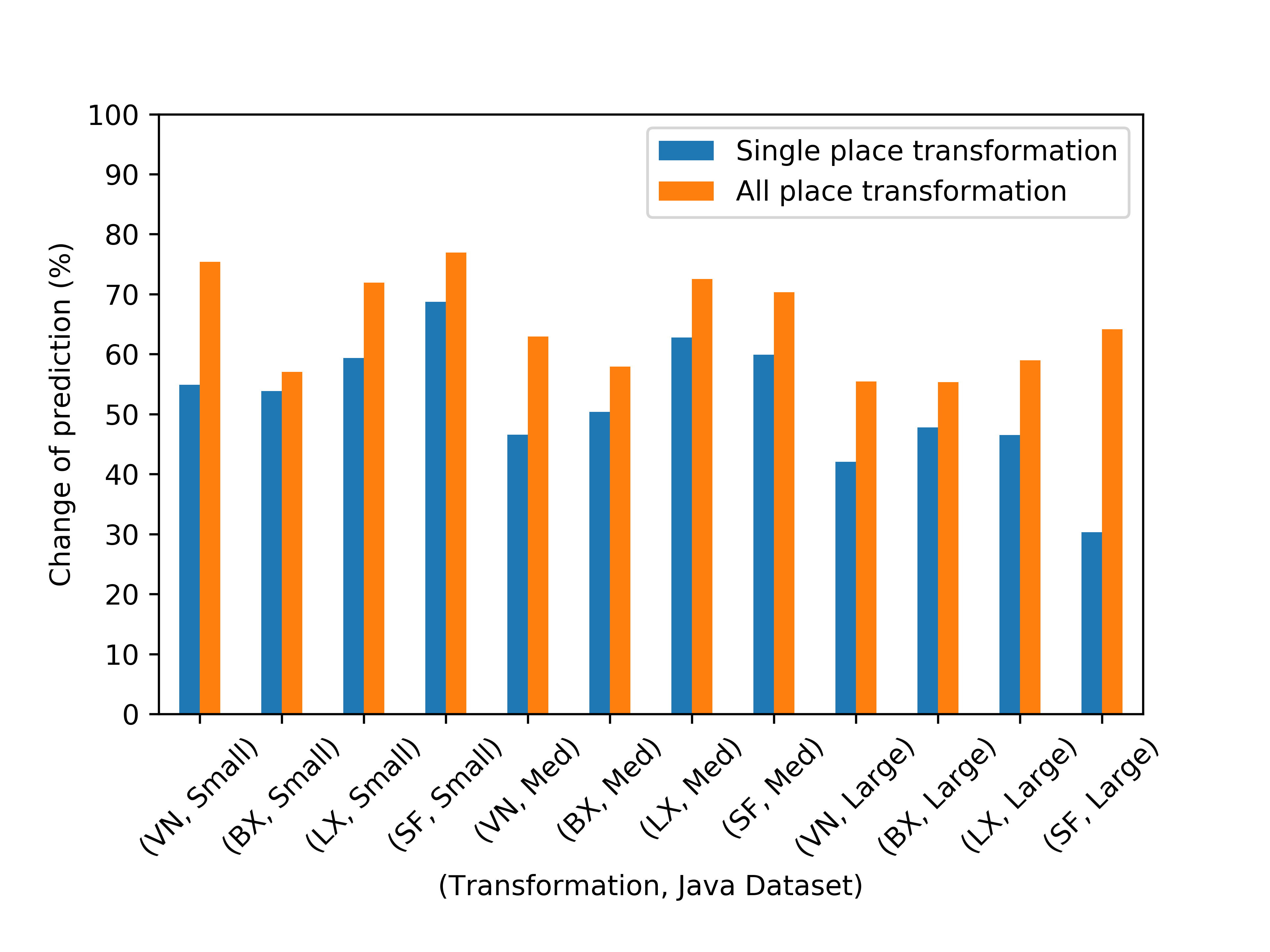}
\end{minipage}%
\begin{minipage}{.5\textwidth}
\caption*{(b) \cts}
\includegraphics[width=0.99\linewidth]{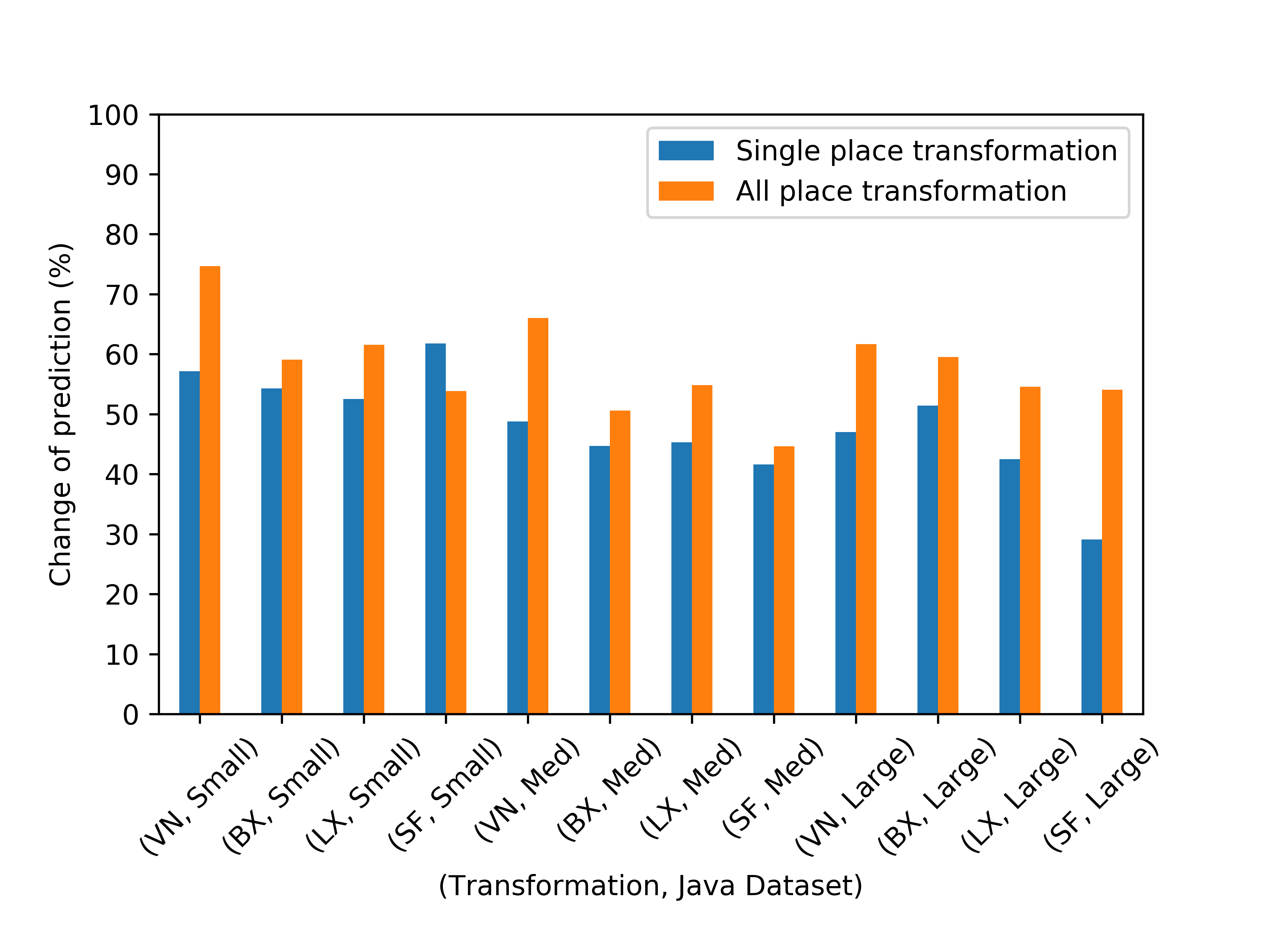}
\end{minipage}
\caption{Change of prediction after transformation in single-place vs all-place.}
\label{fig:single_vs_all}
\end{figure*}

Figure~\ref{fig:single_vs_all} compares the impact of single-place transformation and all-place transformation on the change of prediction in all \npas that we studied.
For the \ctv model, the percentage of prediction change for all-place transformation is higher than the single-place transformation by a good margin for all the cases. 
Similarly, for the \cts model, the prediction change percentage for all-place transformation is higher than the single-place transformation by a good margin except for the case (\SF, \Js).
After a closer examination of \Js dataset and \SF transformation, we observe that the number of transformed methods for all-place is only $13$, which is too low to provide comparative insight.

\observation{All-place transformation is more likely to induce changes in the predictions than sing-place transformations.}

\subsubsection{Correctly predicted methods vs Incorrectly predicted methods}

\begin{figure*}
\noindent \begin{minipage}{.5\textwidth}
\caption*{(a) \ctv}
\includegraphics[width=0.99\linewidth]{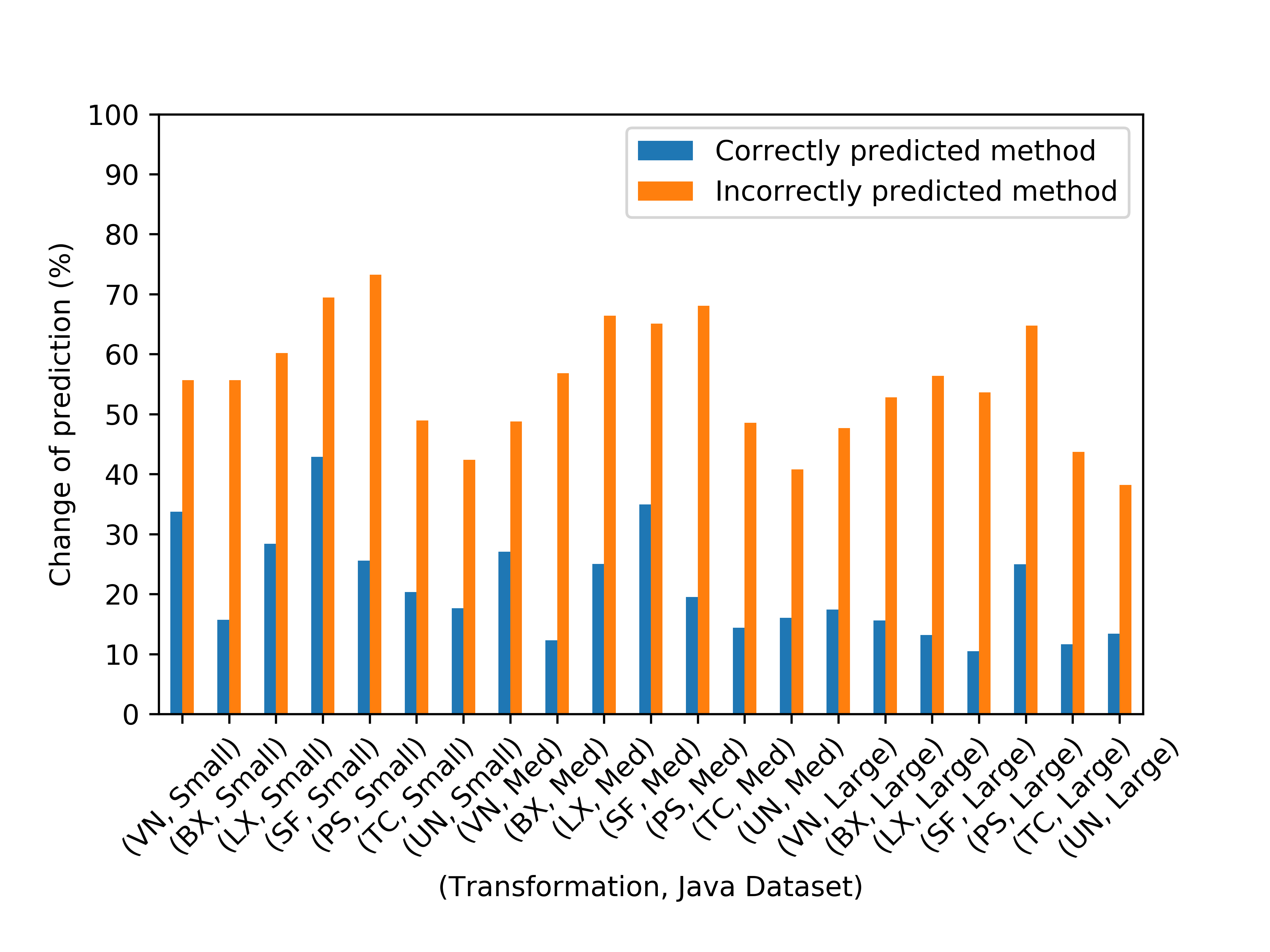}
\end{minipage}%
\begin{minipage}{.5\textwidth}
\caption*{(b) \cts}
\includegraphics[width=0.99\linewidth]{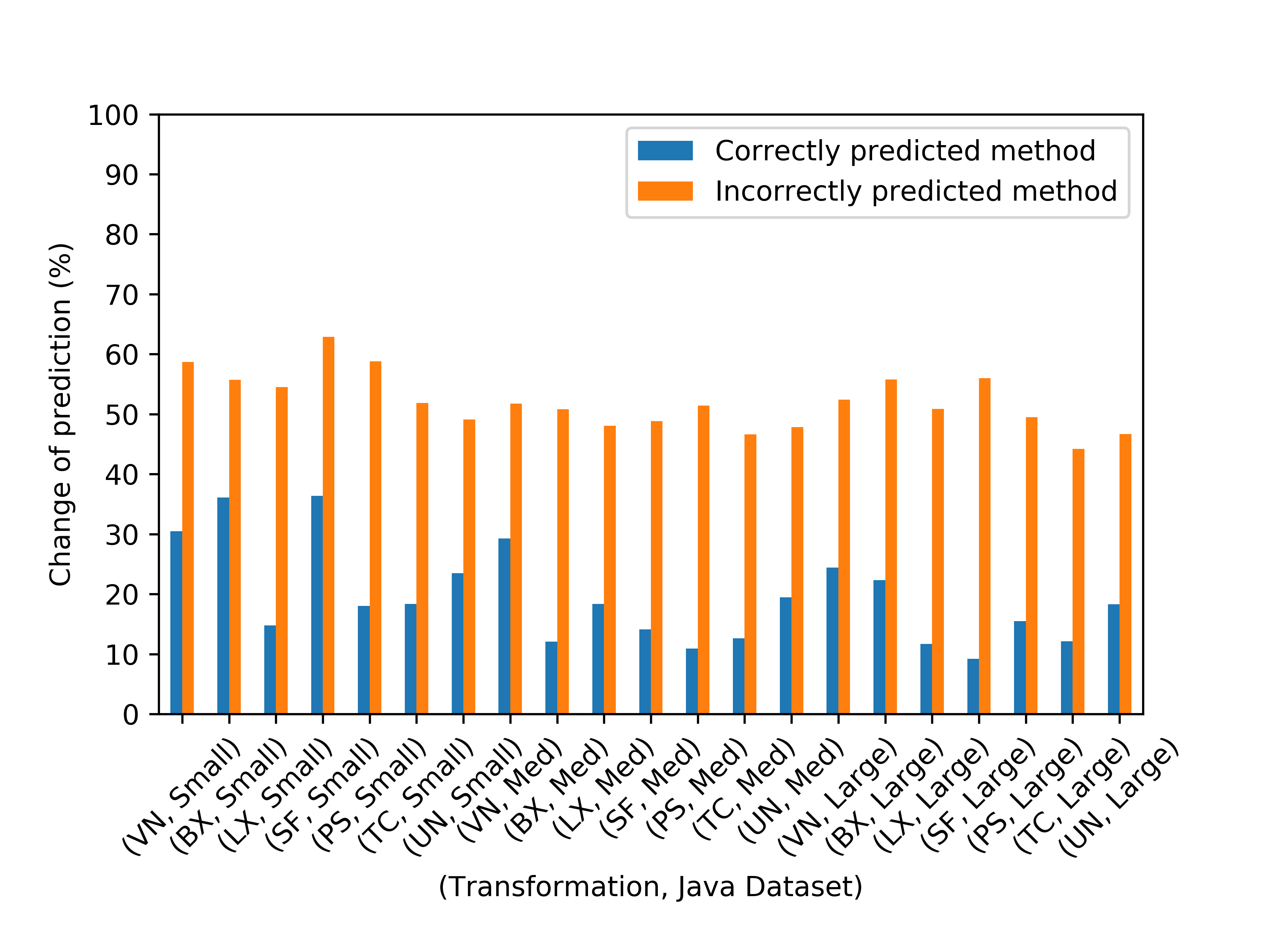}
\end{minipage}
\caption{Change of prediction after transformation on correctly vs incorrectly predicted methods.}
\label{fig:correct_vs_incorrect}
\end{figure*}

Figure~\ref{fig:correct_vs_incorrect} depicts the changes in prediction after prediction on correctly vs. incorrectly predicted methods in all \npas.
In \ctv \npas, the percentage of prediction changes after transformation in the correctly predicted methods ranges from 10.45\% to 42.86\%, while, in the incorrectly predicted methods, a larger portion of transformations, 38.18\% to 73.25\%, change the prediction of \ctv.
In \cts \npas, while the percentages of changes in predictions after transformation on the correctly predicted methods ranges from 9.19\% to  36.36\%, the percentages range from 44.18\% to 62.9\% in the incorrectly predicted methods.
\observation{The changes in prediction happens more frequently in the originally incorrect methods.}

\subsection{RQ3: Impact of Method Length on Generalizability}

\begin{figure*}
\noindent \begin{minipage}{.33\textwidth}
\caption*{(a) \ctv (\Js)}
\includegraphics[width=0.99\linewidth]{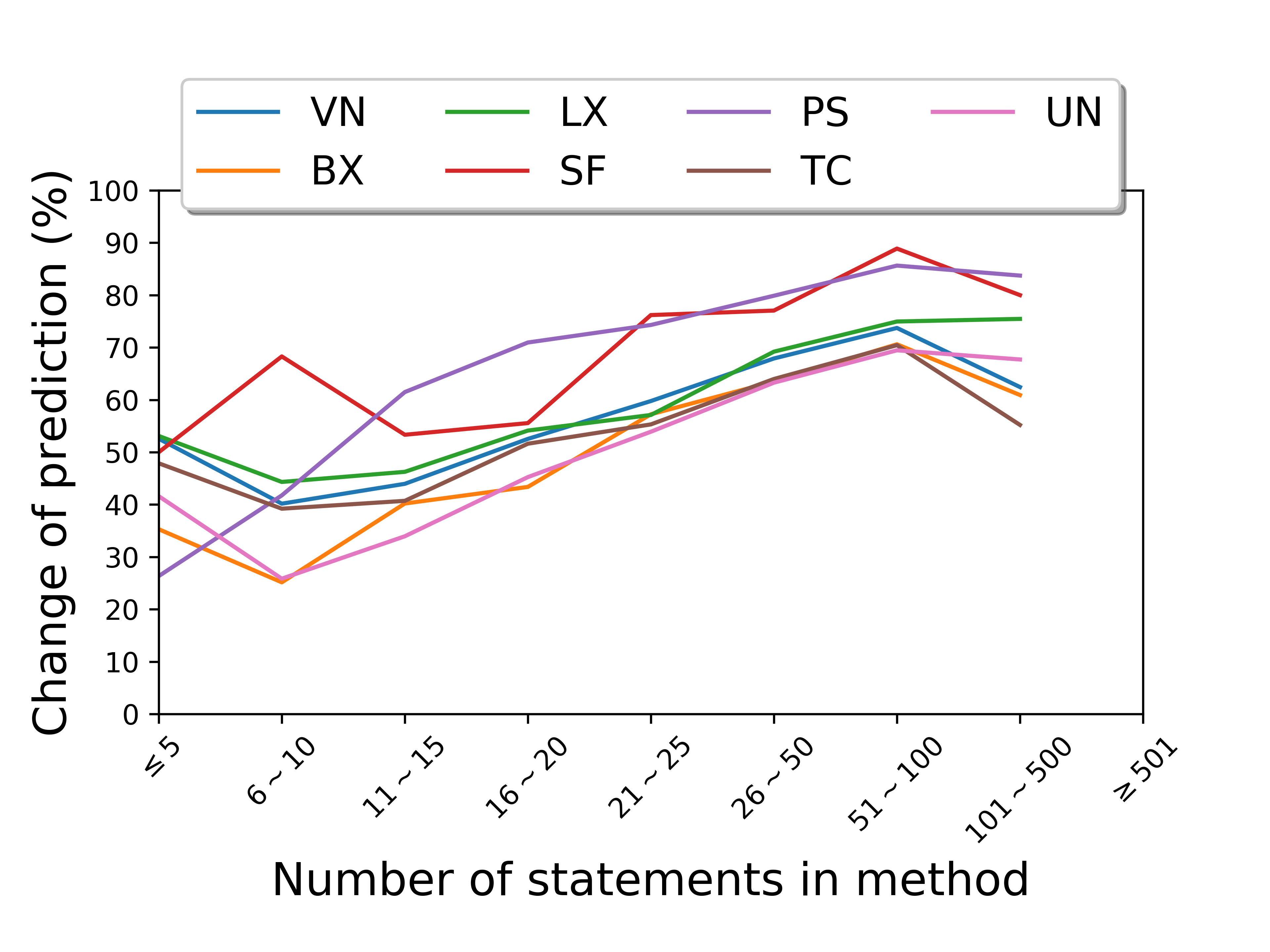}
\end{minipage}%
\begin{minipage}{.33\textwidth}
\caption*{(b) \ctv (\Jm)}
\includegraphics[width=0.99\linewidth]{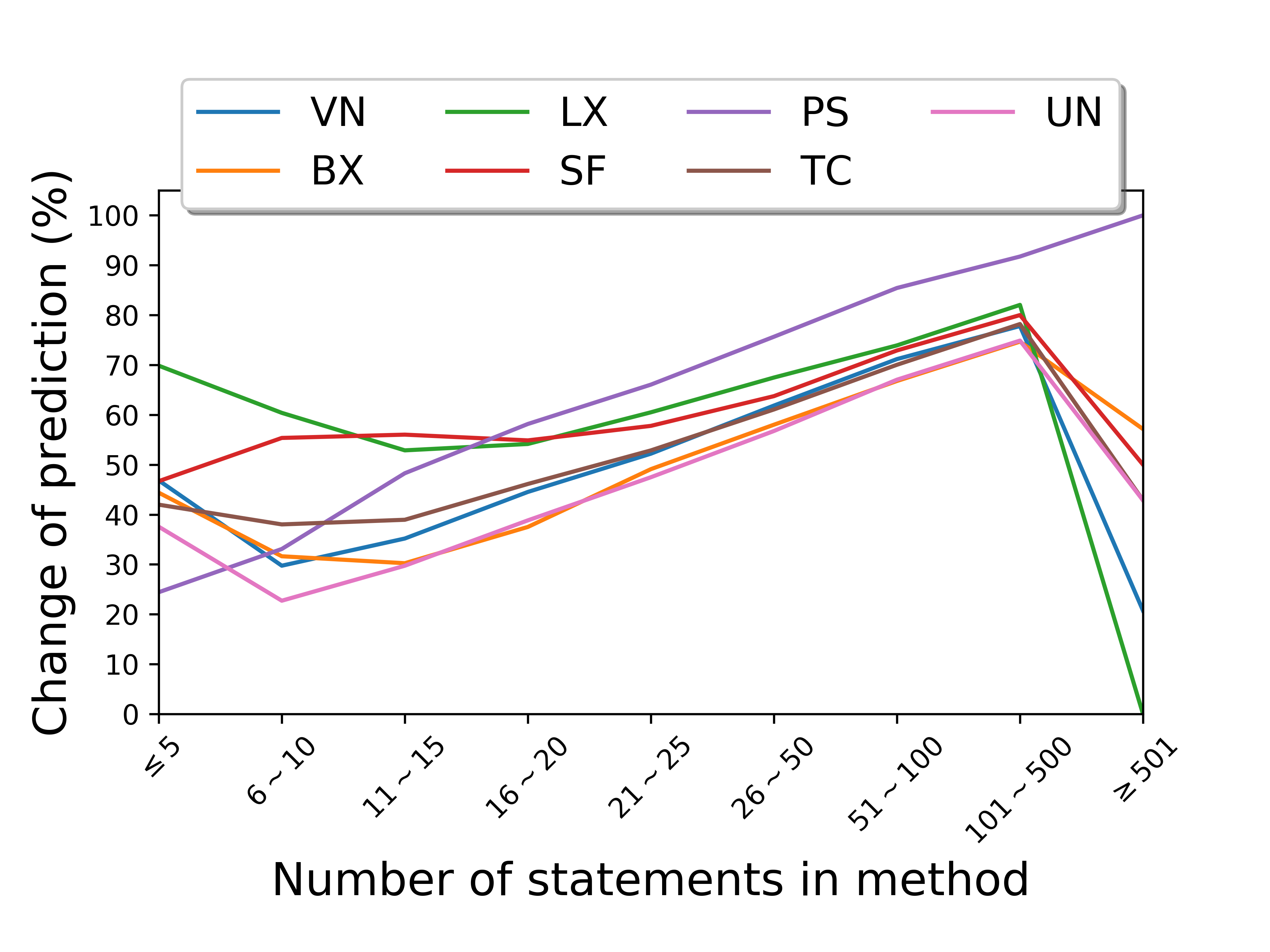}
\end{minipage}%
\begin{minipage}{.33\textwidth}
\caption*{(c) \ctv (\Jl)}
\includegraphics[width=0.99\linewidth]{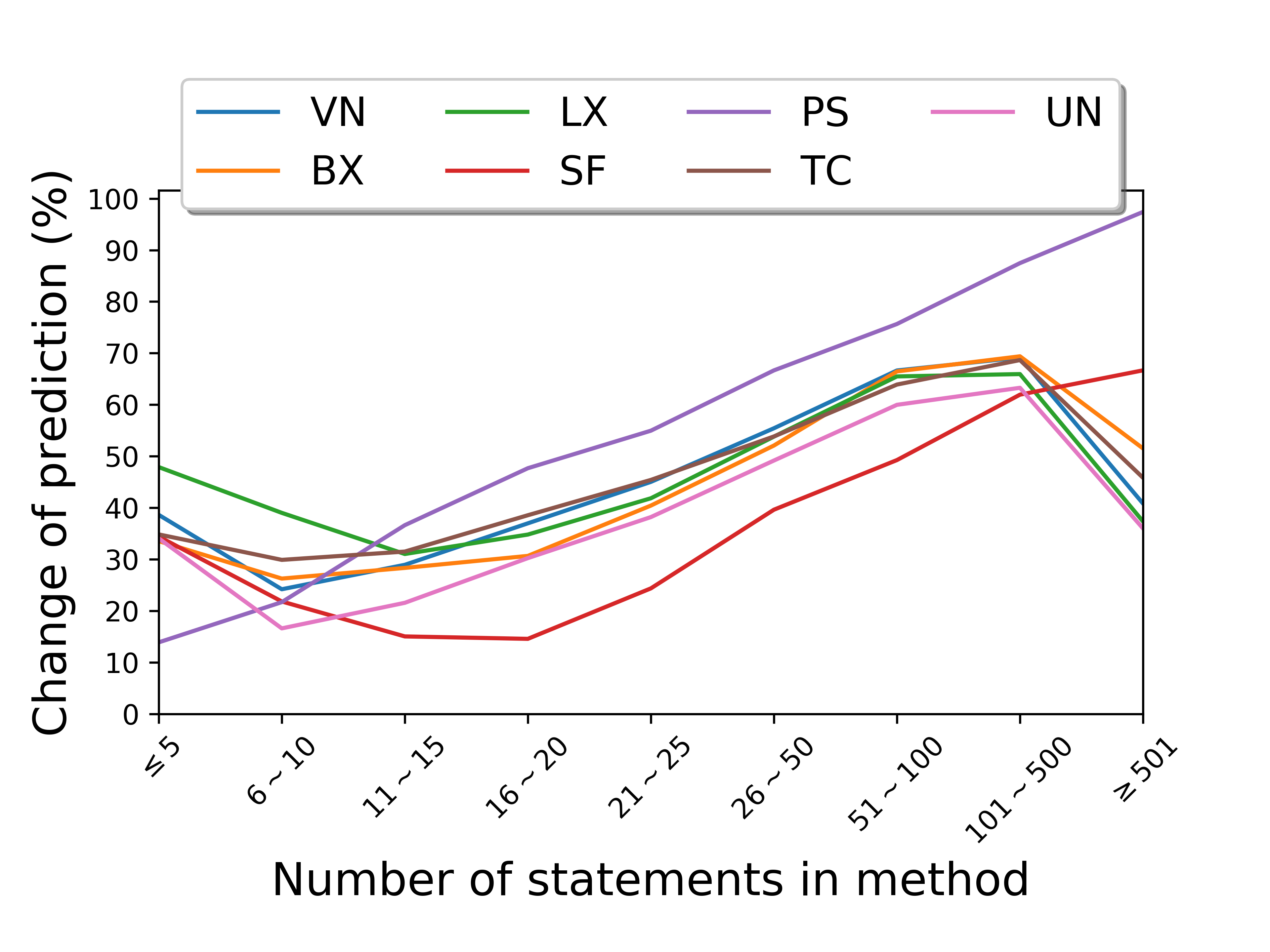}
\end{minipage}
\noindent \begin{minipage}{.33\textwidth}
\caption*{(d) \cts (\Js)}
\includegraphics[width=0.99\linewidth]{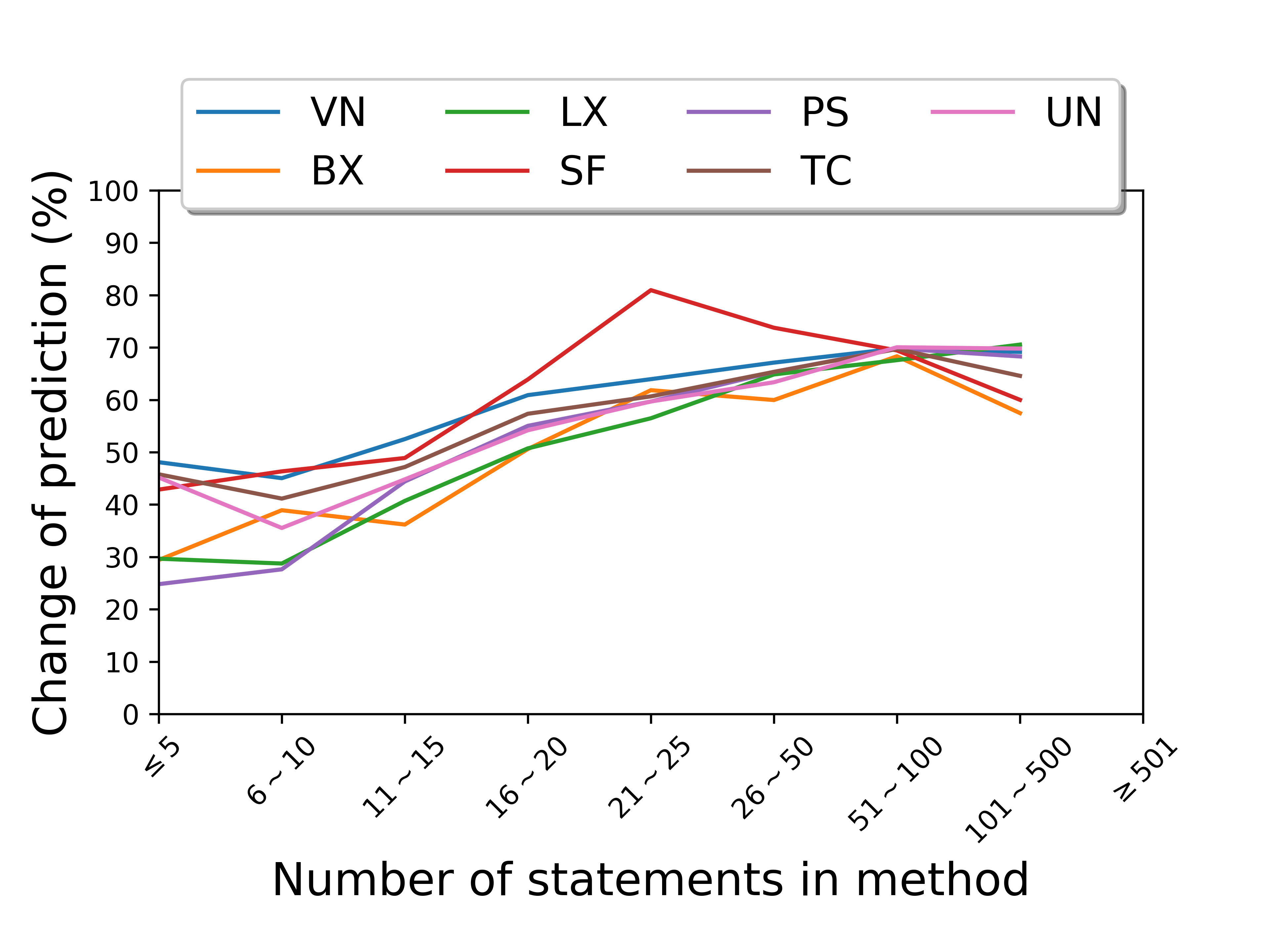}
\end{minipage}%
\begin{minipage}{.33\textwidth}
\caption*{(e) \cts (\Jm)}
\includegraphics[width=0.99\linewidth]{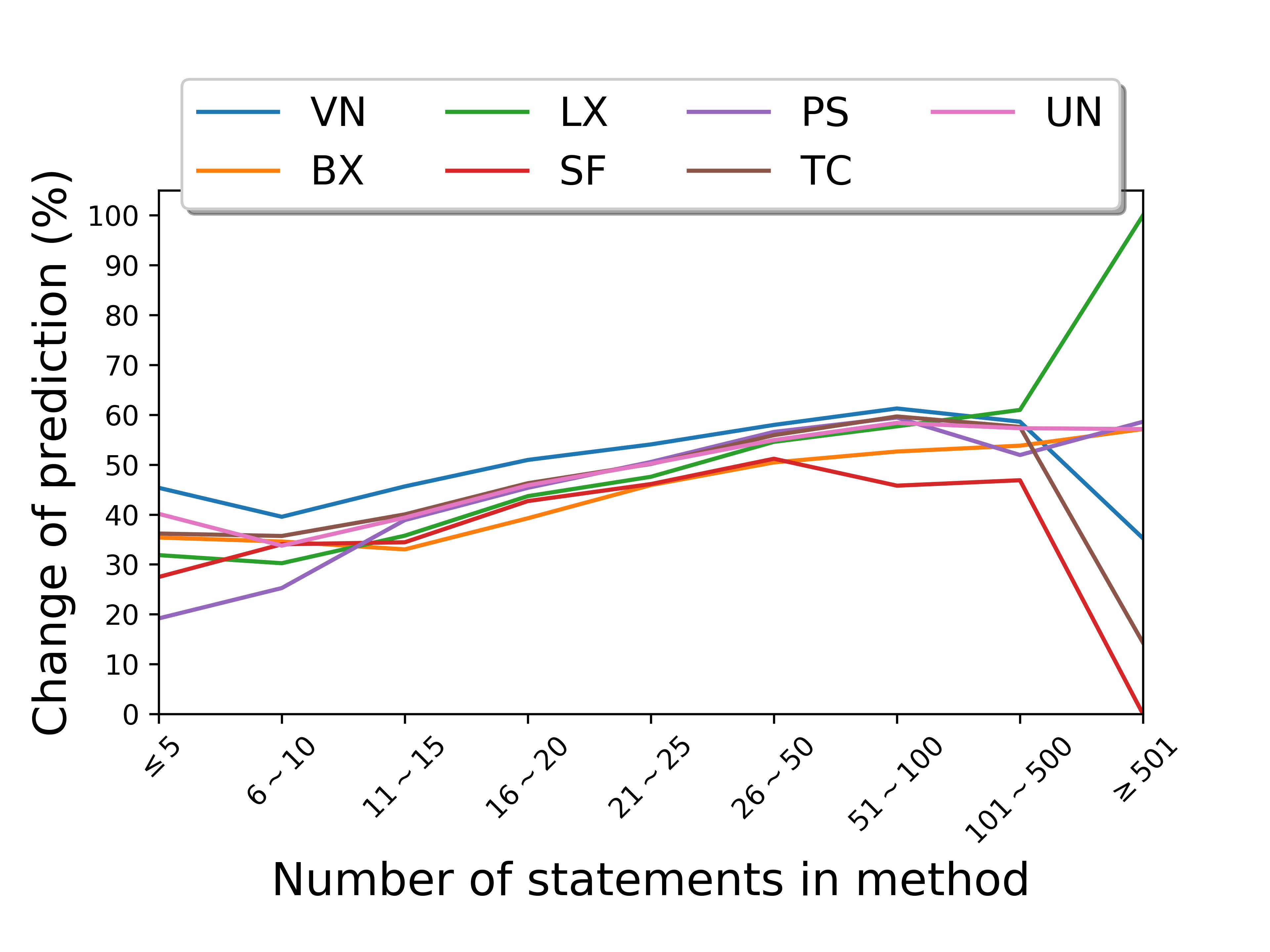}
\end{minipage}%
\begin{minipage}{.33\textwidth}
\caption*{(f) \cts (\Jl)}
\includegraphics[width=0.99\linewidth]{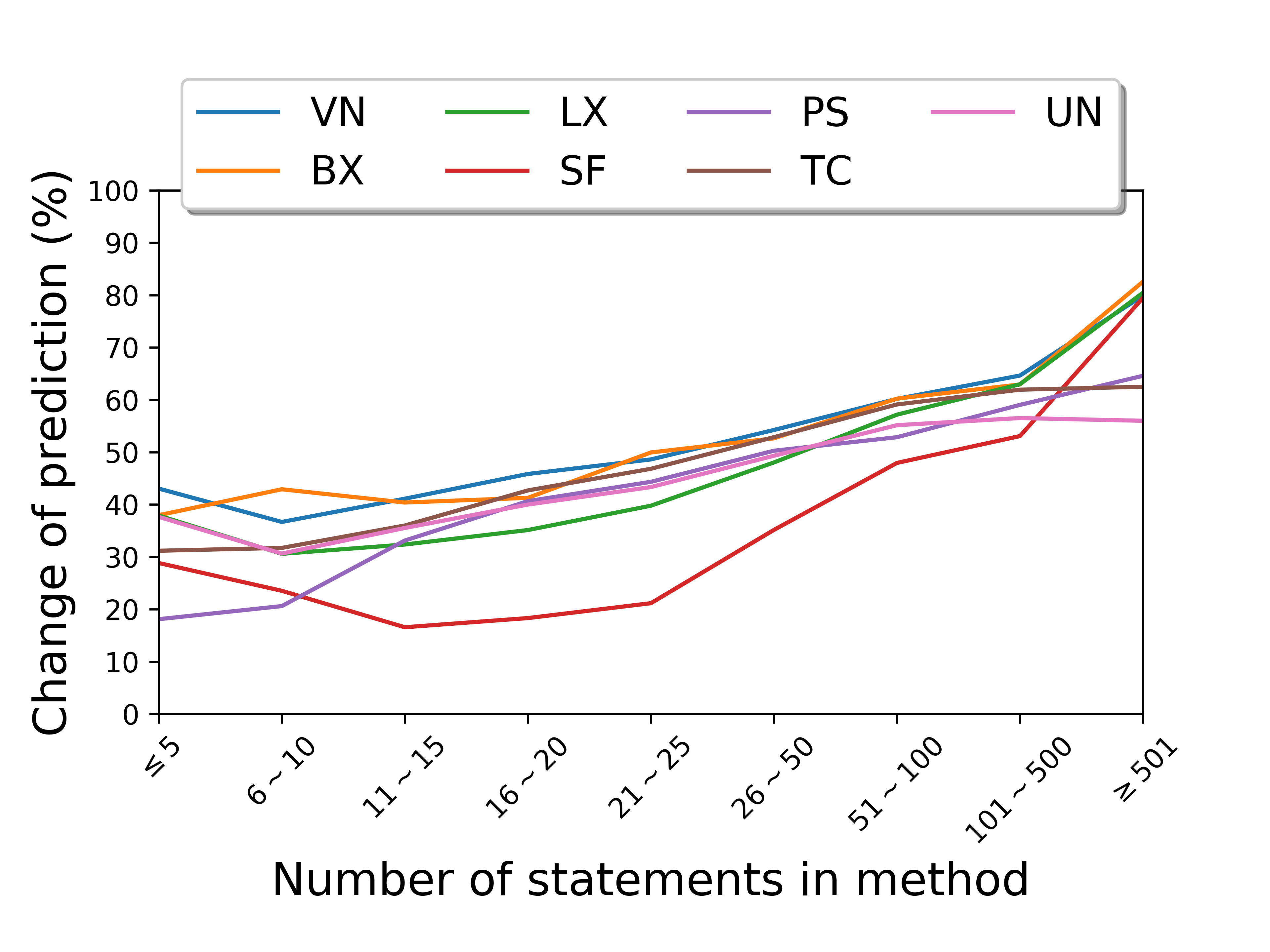}
\end{minipage}
\caption{Change of prediction for the number of statements in method.}
\label{fig:num_of_stmt}
\end{figure*}

An important metric of interest might be the generalizability in terms of the number of statements in the methods.
Figure~\ref{fig:num_of_stmt} depicts the relation between length of methods and percentage of prediction changes.
In the figure, the ``Number of statements in method'' denotes the number of executable lines in the body of methods before transformation.
The trend in the figure suggests that there is a direct correlation between the size of programs and the percentage of predication changes after transformation. Note that there are only a handful of programs larger than 500 lines, therefore their behavior in Figures ~\ref{fig:num_of_stmt}-b and ~\ref{fig:num_of_stmt}-e is an outlier and can be ignored.

For \Js dataset, the prediction change of models for all transformations is more superior when a method has around $100$ statements and there is no method more than $500$ statements.

For \Jm dataset, the prediction change of models for all transformations is more superior when a method exceeds $100$ statements but drops significantly once a method exceeds $500$ statements, except for the \ctv model on \PS transformation and the \cts model on \LX transformation where the prediction change increases even method exceeds $500$ statements.

For \Jl dataset, the prediction change of \ctv model for all transformations is more superior when a method exceeds $100$ statements but drops significantly once a method exceeds $500$ statements, except for the \PS and \SF transformation where the prediction change increases even method exceeds $500$ statements. On the other hand, the prediction change of \cts model for all transformations is increasing whether method exceeds $500$ statements.

As shown in Figure~\ref{fig:num_of_stmt}, most of the cases the models exhibit notable increases in prediction change for all transformations as the number of lines in the program increases.

\observation{There is a direct correlation between the size of methods and their susceptibility of changes in the prediction under transformations.}

\begin{table}
  \begin{center}
    \caption{Detailed percentages of changes in \cts (\Jl).}
    \label{tab:new-detail}
    \begin{tabular}
    {|c|c|c|c|c|c|}\hline
    Transformation & CCP& CIP& WWSP& WCP&WWDP \\ \hline \hline
    
        VN &  14.54& 4.69& 38.42& 2.32& 40.03 \\  \hline
    
        BX  & 10.14& 2.92& 38.42& 2.92& 45.59 \\  \hline
        LX  & 18.87& 2.49& 38.62& 2.3& 37.71 \\  \hline
            SF  & 52.23& 5.28& 18.68& 2.8& 20.99 \\  \hline
            PS  & 14.77& 2.7& 41.7& 2.82& 38.0 \\  \hline
            TC & 24.17& 3.33& 40.47& 2.1& 29.93 \\  \hline
    
        UN  & 26.57& 5.95& 35.99& 1.81& 29.68 \\  \hline

    \end{tabular}%
  \end{center}
\end{table}

Table~\ref{tab:new-detail} shows the full breakdown of percentage of changes after transformations in \cts(\Jl) for all transformed programs.
Note that we only refer to the \cts(\Jl) model in this section for its highest F1 score showed in Table~\ref{table:all_models}.
In this table, CCP, CIP, WWSP, WCP, and WWDO respectively denote the percentage of correct predictions that stay correct, percentage of correct predictions that become incorrect, percentage of wrong labels that stay the same after transformations, percentage of wrong predictions that become correct, and percentage of wrong predictions that change to another wrong prediction after the transformation.  
$\frac{CIP}{CCP+CIP}$ calculates the percent of cases that the \npa's prediction has switched from correct to incorrect. In 9\% to 36\% (average \%18) of cases, the \npa switches from a correct prediction to a wrong one in  \cts(\Jl).
$\frac{WCP}{WWSP+WWDP+WCP}$ calculates switching from a wrong prediction to a correct prediction after transformations. Overall, in less than 3\% of transformations, this switch happens.   

\observation{In less than 3\% of cases, a transformation switches from a wrong prediction to correct prediction.}


\section{Discussion}
In this paper, we study the current state of generalizability in two \npas. Although limited, it provides interesting insights. In this section, we first discuss why neural networks have become a popular, or perhaps the de-facto, tool for processing programs, and what are the implications of using neural networks in processing source code.

Neural networks constitute a powerful class of machine learning models with a large hypothesis class. For instance, a multi-layer feed-forward network is called a universal approximator, meaning, it can essentially represent any function~\cite{hornik1989multilayer}. Unlike traditional learning techniques that require extensive feature engineering and tuning, deep neural networks facilitate representation learning. That is that they are capable of performing feature extraction out of the raw data completely on their own~\cite{lecun2015deep}. Given a sufficiently large dataset,  neural networks with adequate capacity can substantially reduce the burden of feature engineering. 
Availability of a large number of code repositories makes data-driven program analysis a good application for neural networks. However, it is still unknown if neural networks are the best way to process programs ~\cite{Devanbu:FSE:2017} vs. \cite{Sutton:2019:maybe}.

Although the large hypothesis class of neural networks and feature learning make them very appealing to use, the complex models built by neural networks are still too difficult to understand and interpret. Therefore as we apply neural networks in program analysis, we should develop specialized tools and techniques to enhance its interpretability of \npas.

\subsection{Are we there yet?}
Are \npas ready for widespread use in program analysis? 
Our results suggest: not yet.
The models that we experimented are brittle to even very small changes in the AST. A correct prediction of the \npas in 9.19\% to 36.36\% cases could change to an incorrect one. 
Although substantial progress has been made in developing \npas for various program analysis and processing tasks,
the literature lacks techniques to rigorously evaluate the reliability of such techniques. 
The recent line of work by Nghi et al. 
\cite{Nghi:ASE:2019} in interpretability of \npas, Rabin et al. \cite{Rabin:ASE:2019a} in testing them, and Yefet et al. \cite{Yefet:Arxiv:2019} are much needed steps in a right direction.

\subsection{Generalizability vs. Robustness}
There is a substantial line of work on evaluating the robustness of neural networks especially in the domain of vision and pattern recognition~\cite{szegedy2013intriguing}. The key insight in such domains is that small, imperceptible changes in input should not impact the result of output. While this observation can be true for domains such as vision, it might not be directly applicable to the discrete domain of \npas, since some minor changes to a program can drastically change the semantic and behavior of the program.
Quantifying the imperceptibility of source code is our future research goal.  

\subsection{Code Representation}
The performance of models used in \npas, such as ones used in this study, is relatively low compared to the performance of neural models in domains such as natural language understanding~\cite{sarikaya2014application}, text classification~\cite{lai2015recurrent}.
To improve their performance, we would need novel code representations that better capture interesting characteristics of program. 

\section{Related Work}
\Part{Robustness of neural networks}
There is a substantial line of work on robustness of AI systems in general and deep neural networks in particular.~\citet{szegedy2013intriguing} is the first to discover deep neural networks are vulnerable to small perturbations that are imperceptible to human eyes. They developed the L-BFGS method for systematic generation of such adversarial examples.~\citet{goodfellow2014explaining} proposes a more efficient method, called Fast Gradient Sign Method that exploits the linearity of deep neural networks. Many following up works~\cite{kurakin2016adversarial,moosavi2016deepfool,carlini2017towards,dong2018boosting} further demonstrated the severity of the robustness issues with a variety of attacking methods. While aforementioned approaches only apply to models for image classification, new attacks have been proposed that target models in other domains, such as natural language processing~\cite{li2016understanding,jia2017adversarial,zhao2018generating} and graphs~\cite{Dai_Li_Tian_Huang_Wang_Zhu_Song_2018,zugner2018adversarial}.

Automated verification research community has proposed techniques to offer guarantees for robustness of neural networks by adapting bounded model checking~\cite{scheibler2015towards}, abstract interpretation~\cite{gehr2018ai2}, and Satisfiability Modulo Theory~\cite{HX}.


\Part{Models of Code}
Early works directly adopted NLP models to discover textual patterns existed in the source code~\cite{gupta2017deepfix,pu2016sk_p}. Those methods unfortunately do not account for the structural information programs exhibit. Following approaches address this issue by generalizing from the abstract syntax trees~\cite{maddison2014structured,alon2018code2vec,alon2018code2seq}. As Graph Neural Networks (GNN) have been gaining increasing popularity due to its remarkable representation capacity, many works have leveraged GNN to tackle challenging tasks like program repair and bug finding, and obtained quite promising results~\cite{allamanis2017learning,wang2019learning,dinella2020hoppity}. In parallel, Wang~\Etal developed a number of models~\cite{wang2017dynamic,wang2018dynamic,DBLP:journals/corr/abs-1907-02136} that feed off the run time information for enhancing the precision of semantic representation for model inputs.


\section{Conclusion}

In this paper, we perform a large-scale, comprehensive evaluation on the generalizability of code2vec and code2seq.
In particular, we apply semantics preserving program transformations to produce new programs on which we expect models to keep their original predictions. 
We find that such program transformations frequently sway the predictions of both models, indicating the serious generalization issues that could negatively impact the wider applications of deep neural networks in program analysis tasks. 
We believe our work can motivate the future research on training not only accurate but also robust deep models of code.


\bibliographystyle{ACM-Reference-Format}
\bibliography{references}
\end{document}